\documentclass[%
 reprint,
 amsmath,amssymb,
 aps,
]{revtex4-1}

\usepackage{graphicx}
\usepackage{dcolumn}
\usepackage{bm}
\usepackage{epstopdf}

\begin{document}

\preprint{APS/123-QED}

\title{Magneto-electroluminescence of organic heterostructures: \\Analytical theory and spectrally resolved measurements}

\author{Feilong Liu$^1$}
 \email{liux0756@umn.edu} 

\author{Megan R. Kelley$^2$, Scott A. Crooker$^2$, Wanyi Nie$^2$, Aditya D. Mohite$^2$, P. Paul Ruden$^{1,2}$}
\author{Darryl L. Smith$^{1,2}$}
\affiliation{$^1$University of Minnesota, Minneapolis, MN 55455 USA}
\affiliation{$^2$Los Alamos National Laboratory, Los Alamos, NM 87545 USA}

\date{\today}
\begin{abstract}
The effect of a magnetic field on the electroluminescence of organic light emitting devices originates from the hyperfine interaction between the electron/hole polarons and the hydrogen nuclei of the host molecules.  In this paper, we present an analytical theory of magneto-electroluminescence for organic semiconductors.  To be specific, we focus on bilayer heterostructure devices.  In the case we are considering, light generation at the interface of the donor and acceptor layers results from the formation and recombination of exciplexes.  The spin physics is described by a stochastic Liouville equation for the electron/hole spin density matrix.  By finding the steady-state analytical solution using Bloch-Wangsness-Redfield theory, we explore how the singlet/triplet exciplex ratio is affected by the hyperfine interaction strength and by the external magnetic field.  To validate the theory, spectrally-resolved electroluminescence experiments on BPhen/m-MTDATA devices are analyzed.  With increasing emission wavelength, the width of the magnetic field modulation curve of the electroluminescence increases while its depth decreases.  These observations are consistent with the model.  Finally, the analytical theory is extended to account for an additional low-field structure due to the exchange interaction in the weakly bound polaron-pair states. 

\begin{description}

\item[PACS numbers]
81.05.Fb, 85.60.Jb, 78.20.Ls
\end{description}
\end{abstract}

%81.05.Fb Organic semiconductors
%85.60.Jb Light-emitting devices
%78.20.Ls Magneto-optical effects


\maketitle

\section{Introduction}

Organic light emitting devices (OLEDs) are potential candidates for future display technologies due to advantages such as high contrast ratio, light weight, and flexibility.  In addition, the field of spintronics has recently expanded into the organic semiconductor realm because of relatively long spin coherence times, which is critical for applications such as organic spin valves \cite{ZVbook,ZVnat,BKnat,VDnat}.  Exploration of the modulation of OLED light emission by an applied external magnetic field combines these two areas of research \cite{1,2,3,4,5,6,7,8,9,10}.  An increase in electroluminescence of up to 10\% in small magnetic fields has been observed in experiments \cite{1,6,7,9}, and the physics originating from the hyperfine interaction between electron/hole polarons and hydrogen nuclei in the host molecules has begun to be explored \cite{11,12}.  The study of magneto-electroluminescence (MEL) also has potential for the development of organic semiconductor spintronics and for adding insight into the physics of charge carriers in the organic semiconductors and at organic/organic interfaces.  In this work, we present an analytical model for MEL using a spin density matrix approach.  After establishing rate equations for the relevant microscopic processes, we obtain steady-state solutions.  We explore theoretically the competition between the hyperfine interaction, which expedites spin mixing, and the Zeeman effect which tends to suppress it.  We then compare our results with experimental data on BPhen/m-MTDATA heterojunction OLEDs.  

\section{Theory}

\begin{figure}
\centering\includegraphics[width=3.37in]{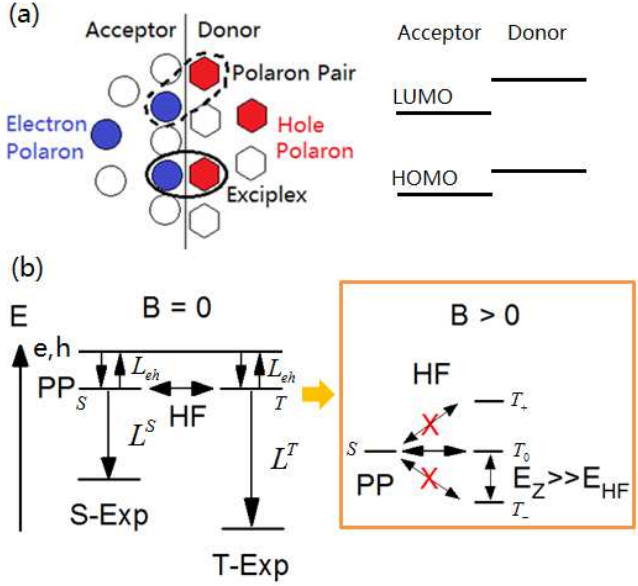}% Here is how to import EPS art
\caption{\label{1} (Color Online) (a) A schematic of the acceptor/donor interface and its band diagram. (b) The energy levels and critical processes in a heterojunction OLED.  An external magnetic field tends to suppress spin mixing in PP states.  HF denotes hyperfine interaction, S-Exp and T-Exp are short for singlet and triplet exciplexes, e,h stands for electron and hole polarons, and other acronyms are explained in the text.}
\end{figure}

A schematic diagram of the donor/acceptor interface of an OLED is shown in Fig. \ref{1}.  An electron (hole) and its host acceptor (donor) molecule form an electron (hole) polaron.  Under forward bias, the electron and hole polarons move towards the acceptor/donor interface, where polaron pairs may form due to their mutual Coulomb attraction.  A polaron pair (PP) is envisioned as a relatively weakly bound state in which the electron and hole polarons reside on different molecules in relatively close but not necessarily immediate proximity.  A PP can relax to a more tightly bound state with lower energy called an exciplex \cite{13,14}.  In the exciplex state the electron/hole polarons reside on acceptor/donor molecules that are near neighbors.  The exciplex may eventually decay radiatively (resulting in light emission) or non-radiatively.  In different systems other processes may occur.  For example, an electron or hole may overcome the interfacial energy barrier and form a PP in one material, followed by relaxation to a bulk exciton and subsequent radiative or non-radiative decay.  The model presented below is generally applicable to both cases, but to be specific, the following discussions are based on the \textquotedblleft interface\textquotedblright processes sketched in Fig. \ref{1}(a).

Assuming non-polarized carrier injection from the contacts, the spins of the electron and hole polarons are randomly oriented with two possible values, up ($\uparrow$) and down ($\downarrow$).  For the PPs, we consider the spin configurations parallel ($\uparrow\uparrow$, $\downarrow\downarrow$) and antiparallel ($\uparrow\downarrow$, $\downarrow\uparrow$) as the basis set.  The first (second) arrow denotes the electron (hole) spin.  As the interaction between the two polarons in the PP state is relatively weak, the exchange coupling between the two spins can be neglected for the moment. (This effect is incorporated into the theory in section IV.)  In the PP state different spin configurations have approximately the same energy at zero magnetic field.  However, in the exciplex state, a strong Coulomb interaction leads to a significant exchange splitting between singlet and triplet states.  As shown in Fig. \ref{1}(b), the exciplex energy levels for singlet and triplet states are therefore different, and the exciplex formation rates from the polaron pair ($L^{S}$ and $L^{T}$) are also different.

The PP state is an important intermediate step, in which spin flips (intersystem crossing) may occur without changing the energy significantly if there is no applied magnetic field.  The most important mechanism for spin flips in the organic semiconductors under consideration is the hyperfine interaction between the polarons and (typically many) hydrogen nuclei in the molecules \cite{9,11,12}.  In this work, the hyperfine interaction is assumed to be isotropic; therefore the MEL is independent of the direction of the magnetic field.  The spin-orbit interaction is negligible due to the light elements (C, H, O) composing the molecules \cite{12}. (Exceptions like molecules containing heavy metal atoms \cite{15,16} are not considered in this work.)  When an external magnetic field is present, the Zeeman effect splits the energy levels of the different PP states.  In general, the mixing of states due to hyperfine interaction is then suppressed, resulting in a magnetic-field dependence of the luminescence (Fig. \ref{1}(b)).  On the other hand, due to the large exchange splitting, the exciplex states do not mix.

The observed luminescence is the result of recombination of a large number of singlet exciplexes generated from their precursor PP states.  A convenient tool with which to describe the relevant ensemble of polaron pair states is a density matrix, which has been employed previously for modeling the magnetoresistance of organic semiconductors \cite{17,18}.  The four dimensional PP spin Hilbert space in this study is spanned by the combination of electron and hole polaron spin states.  The system Hamiltonian includes both the Zeeman ($H_Z$) and the hyperfine ($H_{HF}$) interaction.  We choose the Zeeman interaction as the $0th$ order Hamiltonian and the hyperfine interaction as a perturbation.  The two terms are expressed as \cite{12,19}:

\begin{subequations}
\label{eq1}
\begin{gather}
 H_Z=\sum_{n}Q_n(t)H_Z^n
\\
 H_{HF}=\sum_{n}Q_n(t)H_{HF}^n
\\
 H_Z^n=\frac{g\mu_B}{\hbar}\vec B \cdot (\vec S_h^n + \vec S_e^n)
\\
 H_{HF}^n=\sum_{i_n}\big|\psi_e(r_{i_n})\big|^2a_{i_n}\vec S_e^n \cdot \vec N_{i_n}\nonumber\\
 +\sum_{k_n}\big|\psi_h(r_{k_n})\big|^2a_{k_n}\vec S_h^n \cdot \vec N_{k_n}
\end{gather}
\end{subequations}

Here $n$ labels the various molecular pair sites that can support a polaron pair state, $Q_n(t)$ is unity if the molecular pair $n$ is occupied by a polaron pair at time $t$ and zero otherwise.  $Q_n(t)$ describes the fact that the polaron pair resides on the site $n$ for a finite period of time and the local hyperfine and external magnetic fields interact coherently with the polarons only during that time (properties of $Q_n(t)$ are discussed in the supplemental material \cite{20}).  $g\approx2$ is the electron/hole g-factor \cite{21}, $\mu_B$ is the Bohr magneton, $\vec S_{e,h}^n$ is the (electron, hole) polaron spin on molecular pair $n$, $i_n$ and $k_n$ label the nuclei that interact with the electron and hole spin at the polaron pair site $n$; $|\psi_{e,h}(r_{i_n,k_n})|^2$ is the squared (electron, hole) wavefunction evaluated at the nuclear position; $\vec N$ is the nuclear spin, and $a$ is the hyperfine coupling constant.  $\vec B$ is the external magnetic field.  The polaron pair state can form an exciplex state or dissociate into separated electron and hole polarons.

The time evolution of the density matrix, $\rho$, for the PP ensemble is described by a stochastic Liouville equation \cite{22}:

\begin{equation}
\label{eq2}
\frac{d\rho}{dt}=\frac{i}{\hbar}[\rho, H_Z+H_{HF}]+\frac{\partial\rho}{\partial t}\bigg|_{eh}+\frac{\partial\rho}{\partial t}\bigg|_{EP}
\end{equation}

$\frac{\partial\rho}{\partial t}\big|_{eh}$ is the formation rate of PPs from independent electron and hole polarons and the possible dissociation of the polaron pairs back to independent electron and hole polarons.  Assuming charge conservation and spin randomness, this term can be written as:
\begin{equation}
\label{eq3}
 \frac{\partial\rho}{\partial t}\bigg|_{eh}=RI-L_{eh}\rho \tag{3a}
\end{equation}

$R$ is the rate constant for forming polaron pairs from independent electrons and holes, $I$ is the identity operator and $L_{eh}$ is the dissociation rate constant for polaron pairs.  The dissociation rate for polaron pairs is assumed to be independent of polaron spin.  $\frac{\partial\rho}{\partial t}\big|_{EP}$ describes the rate of exciplex formation from PPs.  It is proportional to the PP density, and can be written as \cite{11}:

\begin{equation}
 \frac{\partial\rho}{\partial t}\bigg|_{EP}=-\frac{1}{2}(\Lambda\rho+\rho\Lambda) \tag{3b}
\end{equation}

$\Lambda=\sum_{\lambda}L^{\lambda}|\lambda\rangle\langle\lambda|$ is a projection operator.  $\lambda=S$, $T_0$, $T_+$, or $T_-$ labels four exciplex states.  The singlet state ($S$) and triplet states ($T_0$, $T_+$, $T_-$) are defined as $S=(\uparrow\downarrow-\downarrow\uparrow)/\sqrt{2}$, $T_0=(\uparrow\downarrow+\downarrow\uparrow)/\sqrt{2}$, $T_+=\uparrow\uparrow$, and $T_-=\downarrow\downarrow$.  $L^S$ and $L^T$ (same for three triplet states \cite{11}) are the singlet and triplet exciplex formation rate constants. It is convenient to define the rate constants $K_{S,T}=L_{eh}+L^{S,T}$.  No magnetic field effects on the electroluminescence can arise if $K_S=K_T$, consequently, these quantities cannot be completely dominated by $L_{eh}$. 

\begin{widetext}
\setcounter{equation}{3}
\begin{subequations}
\label{eq4}
\begin{gather}
\rho_{22}=\rho_{33}=\frac{2[K_T+2(J_O^h+J_O^e)]}{(K_S+3K_T)(J_O^h+J_O^e)+K_T(K_S+K_T)-\frac{(K_S-K_T)^2(J_O^h+J_O^e+K_T)}{2(J_O^h+J_O^e)+4(J_S^h+J_S^e)+K_S+K_T}} R
\\
\rho_{23}=\rho_{32}=\frac{K_S-K_T}{2(J_O^h+J_O^e)+4(J_S^h+J_S^e)+K_S+K_T}\rho_{22}
\\
\rho_{11}=\rho_{44}=\frac{R+(J_O^h+J_O^e)\rho_{22}}{J_O^h+J_O^e+K_T}
\end{gather}
\end{subequations}
\end{widetext}

To find the steady state solution for $\rho$, Bloch-Wangsness-Redfield theory is employed \cite{23,24,25,26}.  After lengthy but straight-forward derivation \cite{20}, analytical results for the $\rho$ matrix elements are obtained in Eq. (\ref{4}).  Subscripts 1 to 4 denote spin configurations: $1=\uparrow\uparrow$, $2=\uparrow\downarrow$, $3=\downarrow\uparrow$, and $4=\downarrow\downarrow$.  All other matrix elements are equal to zero in a steady state.  The $J$ terms describe rates of spin mixing, which originates from spin correlation between states at time $t$ and $t+\tau$. (We use $J$ as a generic symbol for $J_O^e$, $J_O^h$, $J_S^e$, and $J_S^h$.) During the time interval $\tau$, the PP experiences random perturbation due to the hyperfine interaction because the electron and hole polarons interact with different nuclei as they hop from molecule to molecule.  The $J$ terms in general can be expressed as \cite{20}:

\begin{equation}
\label{eq5}
J=\frac{\alpha E_{HF}^2}{\hbar^2}\int f(|\tau|/\tau_0)e^{i\tau\Delta E/\hbar}d\tau
\end{equation}

Here $E_{HF}=g\mu_B B_{HF}$ defines the energy scale of the hyperfine interaction, and $\Delta E$ is the Zeeman energy difference between the initial and final states.  (In principle, the dynamics of the nuclear spins could also contribute to the time dependence of the correlation function $f$, but nuclear spin dynamics is slow compared to the polaron hopping times and therefore is not the important consideration.)  For the $J$ terms in Eq. (\ref{eq4}), the superscripts denote the electron ($e$) and hole ($h$) polarons, the subscripts $O$ and $S$ indicate whether a spin flip occurs ($Opposite$ $spin$, in this case $\Delta E=g\mu_B B$) or not ($Same$ $spin$, in that case $\Delta E=0$) during the time interval $\tau$.  The prefactor $\alpha$ equals $2/3$ for the opposite spin case and  $1/3$ for the same spin case, which results directly from the statistical average of off-diagonal ($x$, $y$) and diagonal ($z$) terms in the Pauli matrices.

The function $f(|\tau|/\tau_0)$ in Eq. (\ref{eq5}) is the correlation function.  It is even and monotonically decreases with $\tau$, because under random perturbations the final state gradually loses its relationship to the initial state.  $\tau_0$ describes the relevant time scale for this process.  Specific forms of $f(|\tau|/\tau_0)$ are discussed in the supplemental material \cite{20}.  Two forms for the correlation functions are considered:  

\begin{subequations}
\label{eq6}
\begin{gather}
Type\: I: F\{f(|\tau|/\tau_0)\}=\frac{2\tau_0}{1+\tau_0^2\omega^2}
\\
Type\: II: F\{f(|\tau|/\tau_0)\}=\frac{2\tau_0}{1+\tau_0|\omega|}
\end{gather}
\end{subequations}

Here $F\{f(|\tau|/\tau_0)\}$ is the Fourier transform of $f(|\tau|/\tau_0)$.  The $type$ $I$ function corresponds to the assumption of a single relaxation time, $\tau_0$ \cite{23}.  The $type$ $II$ function is consistent with the  $1/frequency$ ($=2\pi/\omega$) noise power spectrum that is frequently observed experimentally \cite{27,28}.  It can be the result of a range of relaxation times determining the time-decay of the correlation function \cite{29}.  Combining Eqs. (\ref{eq5}) and (\ref{eq6}) the correlation terms can be written explicitly as:

\begin{widetext}
\begin{subequations}
\label{eq7}
\begin{gather}
Type\: I: J_O^{(e,h)}=\frac{4E_{HF(e,h)}^2}{3\hbar^2}\frac{\tau_{0(e,h)}}{1+g^2\mu_B^2\tau_{0(e,h)}^2B^2/\hbar^2},\: J_S^{(e,h)}=\frac{2E_{HF(e,h)}^2}{3\hbar^2}\tau_{0(e,h)}
\\
Type\: II: J_O^{(e,h)}=\frac{4E_{HF(e,h)}^2}{3\hbar^2}\frac{\tau_{0(e,h)}}{1+g\mu_B\tau_{0(e,h)}|B|/\hbar},\: J_S^{(e,h)}=\frac{2E_{HF(e,h)}^2}{3\hbar^2}\tau_{0(e,h)}
\end{gather}
\end{subequations}
\end{widetext}

The four exciplex formation rates are obtained by combining the PP density matrix elements with corresponding formation rate constants:
  
\begin{subequations}
\label{eq8}
\begin{gather}
\chi_S=L^S(\rho_{22}-\rho_{23})
\\
\chi_{T0}=L^T(\rho_{22}+\rho_{23})
\\
\chi_{T+}=L^T\rho_{11}
\\
\chi_{T-}=L^T\rho_{44}
\end{gather}
\end{subequations}

The singlet exciplex formation rate fraction is obtained as:

\begin{equation}
\label{eq9}
Frac(\chi_S)=\frac{\chi_S}{\chi_S+\chi_{T0}+\chi_{T+}+\chi_{T-}}
\end{equation}

The denominator in this expression is independent of the magnetic field, and, to the extent that dissociation of polaron pairs into independent polarons is negligible, equal to $4R$.

The magnetic field effect (MFE) on the singlet exciplex density formed may be defined as:

\begin{equation}
\label{eq10}
MFE(B)=\frac{\chi_S(B)-\chi_S(B=0)}{\chi_S(B=0)}
\end{equation}

After formation, the singlet/triplet exciplex states may decay radiatively/nonradiatively with different lifetimes.

It has been suggested in recent work that the singlet-triplet exciplex splitting at a particular organic/organic interface may be relatively small, thus allowing for thermally-activated intersystem crossing \cite{R1}.  These processes may alter the singlet/triplet exciplex ratio and therefore affect the luminescence.  Assuming a fraction $0<P<1$ of the triplet exciplexes transform into singlets, the overall singlet exciplex formation rate is given by:

\begin{align}
\label{eqr1}
\chi_S' &= \chi_S+(\chi_{T0}+\chi_{T+}+\chi_{T-})P \nonumber\\
&= \chi_S(1-P)+4RP
\end{align}

The singlet-triplet exciplex splitting in the work cited above \cite{R1} was estimated to be $50$ meV.  Hence, for the magnetic field range of interest ($\sim100$ mT, corresponding to $\sim10$ $\mu$eV), $P$ is expected to be independent of the magnetic field.  Therefore,

\begin{align}
\label{eqr2}
MFE'(B) &= \frac{\chi_S'(B)-\chi_S'(0)}{\chi_S'(0)} \nonumber\\
&=[\chi_S(B)-\chi_S(0)]\frac{1-P}{\chi_S'(0)} \nonumber\\
&= MFE(B)\frac{\chi_S(0)}{\chi_S'(0)}(1-P)
\end{align}

The result shows that shape of the $MFE$'$(B)$ curve is identical to that of the $MFE(B)$ curve, the two differing only by a constant scaling factor.

\section{Results and Discussion}

In order to use the theoretical model for MEL calculations, a rough estimate of the parameters is required.  It is useful to introduce an effective (Overhauser) magnetic field that characterizes the strength of the hyperfine interaction described by $H_{HF}\sim g\mu_B B_{HF}$.  Prior literature suggests that this field is on the order of several millitesla for organic molecules \cite{11,12}, and a number of $5$ mT is used in the following calculations.  A first estimate of the correlation time scale, $\tau_0$, is $2$ ns, and we assume it to be the same for electron and hole polarons \cite{30,31}.  The singlet exciplex formation rate constant is taken to be $(0.6$ ns$)^{-1}$ \cite{32,33,34}.

\begin{figure}
\centering\includegraphics[width=3.37in]{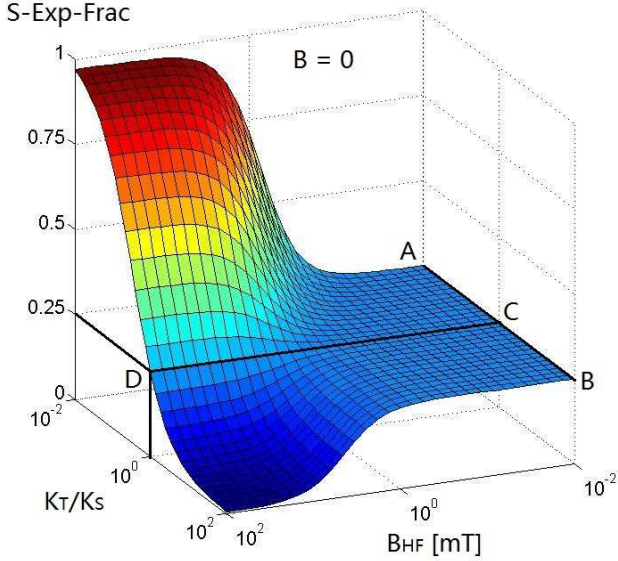}% Here is how to import EPS art
\caption{\label{2} (Color Online) Singlet exciplex fraction plotted as functions of the triplet/singlet formation rate and the strength of the hyperfine interaction.  The external magnetic field is zero.  The parameters are shown in the text.}
\end{figure}

In Fig. \ref{2}, the singlet exciplex fraction is plotted as functions of the ratio $K_T/K_S$ and $B_{HF}$.  For fraction plots in this work, we assume $K_T/K_S=L^T/L^S$ for simplicity.  The external magnetic field is zero.  The results are consistent with numerical simulations shown in Fig. \ref{2}(a) of Ref. \cite{11}.  When the hyperfine interaction is negligible (from point $A$ to point $B$), there is no mechanism for spin perturbation hence the singlet/triplet ratio is constant and equal to $1/3$.  In this case, exciplex formation dominates over spin mixing.  When $K_T/K_S=1$ (from point $C$ to point $D$), the singlet/triplet ratio also maintains the value of $1/3$ regardless of the hyperfine field strength.  In this case, the model does not distinguish between singlet and triplet exciplex states because their formation rate constants are equal.  When the hyperfine interaction is strong and $K_T$ is not equal to $K_S$, the singlet/triplet exciplex ratio is determined by $K_T/K_S$.  In that case the four PP spin states are sufficiently mixed by the hyperfine interaction before exciplex formation can occur.  An external magnetic field suppresses the spin mixing due to hyperfine interaction, hence $K_T/K_S>1$ is required for a positive MEL (Increased electroluminescence with increasing magnitude of the magnetic field).  This study provides some physical insight into the statistical 25\% limit often cited for OLED efficiency.

\begin{figure}
\centering\includegraphics[width=3.37in]{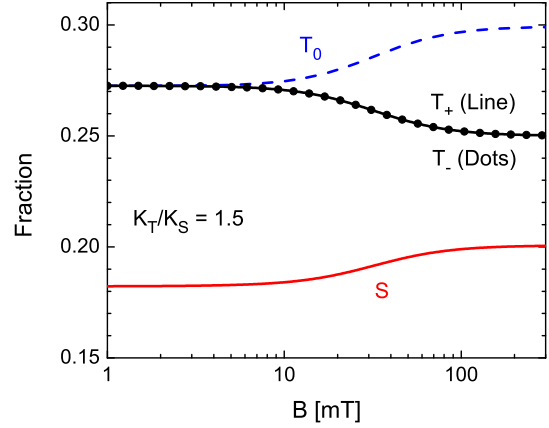}% Here is how to import EPS art
\caption{\label{3} (Color Online) Fraction of singlet and triplet exciplex states as a function of an external magnetic field.  The hyperfine interaction is set to a large value ($50$ mT) to ensure sufficient spin mixing at zero magnetic field.  The correlation function used is $type$ $I$.  Other parameters are the same as in Fig. \ref{2}.}
\end{figure}

The singlet and triplet exciplex fraction as a function of the external magnetic field is shown in Fig. \ref{3}.  Here $K_T/K_S=1.5$.  The hyperfine field $B_{HF}$ is set large enough ($50$ mT) for sufficient PP spin mixing to occur in the absence of an applied magnetic field.  We choose the correlation function in the calculation to be of $type$ $I$.  At zero magnetic field, the singlet and triplet PP spin states (the definitions are the same as for the corresponding exciplex states) have the same energy.  The strong hyperfine interaction leads to substantial spin mixing among all four states.  Their fractions are determined by  $\chi_{T0}=\chi_{T+}=\chi_{T-}=\chi_T$, $\chi_T/\chi_S=K_T/K_S$, and $\chi_S+3\chi_T=1$.  As the magnetic field increases, the Zeeman effect splits the energy degeneracy, and the effect of the hyperfine interaction is gradually suppressed.  When the external magnetic field is strong, the energies of $T_+$ and $T_-$ states are very different from that of the $T_0$ and $S$ states, hence spin flips are suppressed and the exciplex fractions are determined simply by the number of possible states, i.e. both are equal to $1/4$.  On the other hand, $T_0$ and $S$ states are still at the same energy level.  The strong hyperfine interaction determines their fraction through $\chi_{T0}/\chi_S=K_T/K_S$ and $\chi_S+\chi_{T0}=0.5$.  In this calculation $T_+$ and $T_-$ states are symmetric therefore their fractions are always equal. (More discussion of this is presented in section IV.)

\begin{figure} [b]
\centering\includegraphics[width=3.37in]{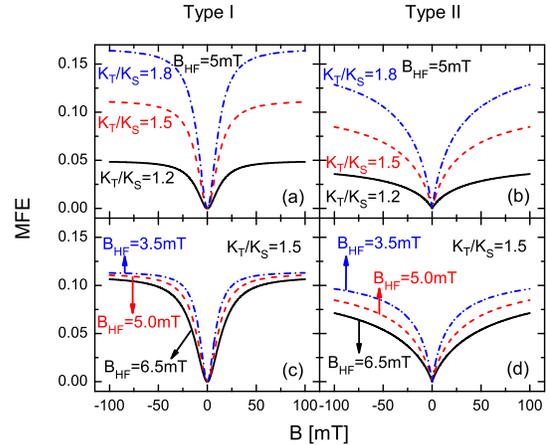}% Here is how to import EPS art
\caption{\label{4} (Color Online) Calculated MFE as a function of the magnetic field, varying three parameters: the hyperfine interaction strength $B_{HF}$, the ratio of triplet/singlet exciplex formation rate $K_T/K_S$, and the type of correlation function.  MFE is defined in Eq. (\ref{eq10}).  The thermally-activated intersystem crossing is assumed to be zero.}
\end{figure}

The MFE curves calculated from Eq. (\ref{eq10}) are plotted as a function of the external magnetic field in Fig. \ref{4}.  The parameters varying are the hyperfine interaction strength $B_{HF}$, the ratio of triplet/singlet exciplex formation rates $K_T/K_S$, and the type of correlation function.  The values used in the calculation are shown in the plots.  The results show that the ‘$depth$’ of the MFE curve is primarily determined by $K_T/K_S$ and the ‘$width$’ is determined by $B_{HF}$.  The difference between the two types of correlation functions is the quadratic or linear dependence on the magnetic field.  The MFE curves with $type$ $I$ function saturate faster than those with $type$ $II$ function as the magnitude of the field increases.  As outlined in the discussion of Eqs. (\ref{eqr1}) and (\ref{eqr2}), the amplitude of the magnetic field modulation of the luminescence may also be affected by intersystem crossing between the exciplex states.  Here we fix $P=0$ and vary the $K_T/K_S$ ratio to fit the measurements, but equally good fits can be achieved fixing that ratio and varying $P$.

\begin{figure} 
\centering\includegraphics[width=3.2in]{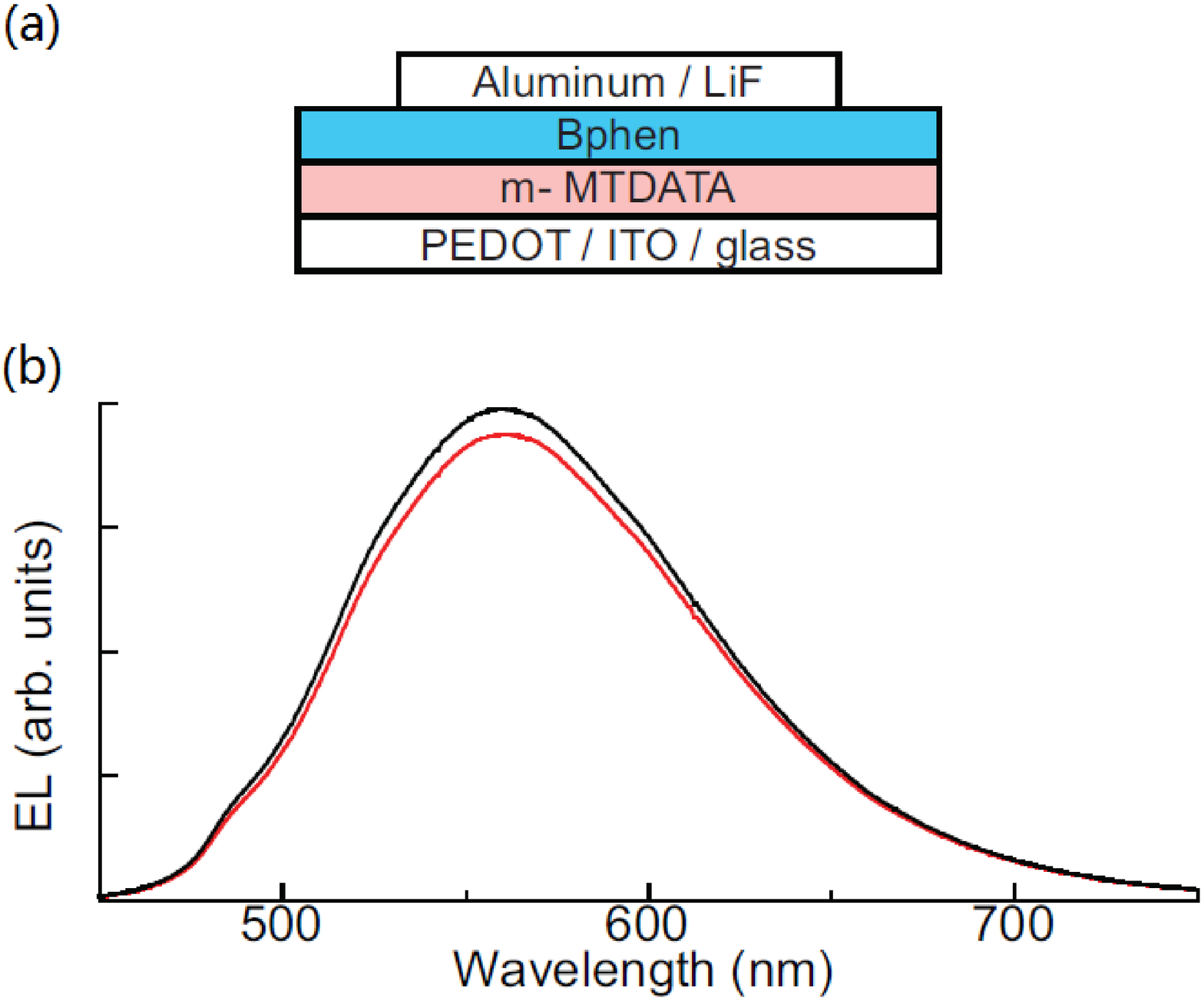}% Here is how to import EPS art
\caption{\label{5} (Color Online) (a) Schematic device structure in experiment.  (b) Measured electroluminescence spectra when the external magnetic field is $0$ (Red) or $100$ mT (black). }
\end{figure}

In the following, the model developed above is applied to the heterojunction OLED structure shown in Fig. \ref{5}(a).  The Al/LiF layer is the cathode and the PEDOT/ITO/glass layer is the anode.  BPhen is the acceptor (electron transport layer) and m-MTDATA is the donor (hole transport layer).  Under forward bias, light emission occurs due to singlet exciplex recombination at the BPhen/m-MTDATA interface.  The measured electroluminescence spectra for $B=0$ (red) and $B=100$ mT (black) are shown in Fig. \ref{5}(b).  More experimental details can be found in a recent paper \cite{35}.

\begin{figure}
\centering\includegraphics[width=3.37in]{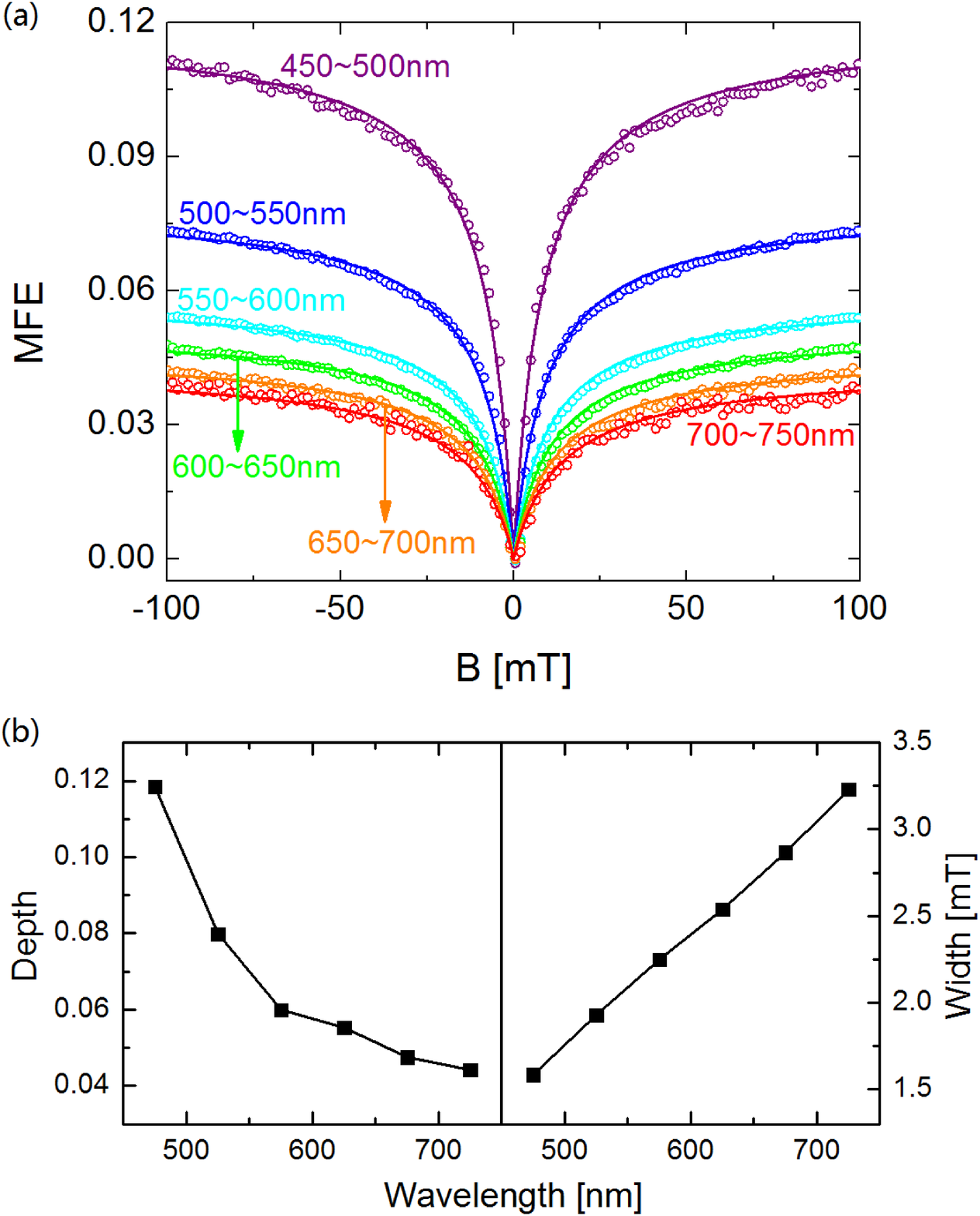}% Here is how to import EPS art
\caption{\label{6} (Color Online) (a) Dots: Measured electroluminescence as a function of the magnetic field.  MFE is defined in Eq. (\ref{eq10}).  Lines: Corresponding model calculation results using a $type$ $II$ correlation function.  The parameters are shown in Table I. (b) The depth and width of the MFE curves as a function of the wavelength.}
\end{figure}

The measured MEL data are normalized using Eq. (\ref{eq10}) and plotted in open circles as a function of the external magnetic field in Fig. \ref{6}(a).  Each curve represents a part of the luminescence spectrum integrated over the wavelength range indicated.  The depth and width of the MFE curves are shown in Fig. \ref{6}(b).  Depth is defined as the difference between the maximum and minimum values of the curve.  The width is taken to be the half-amplitude width.  It is observed that as the wavelength increases, the depth decreases while the width increases.  The different MFE behaviors for different wavelength ranges originate from variations of the interfacial environment.  The molecules at the interface are subject to randomly varying steric interactions with their immediate environment.  Consequently, exciplexes and PP states vary locally in spatial extent and energy, giving rise to the relatively broad spectrum observed.  Generally, the number of hydrogen nuclei that interact with an electron (or hole) polaron is on the order of the number of hydrogen nuclei in the host molecule.  However, due to the steric complexity at the interface, the wavefunctions of an electron (or hole) polaron may vary locally in its spatial extent.  Therefore the number of relevant hydrogen nuclei may vary, resulting in a variation of the hyperfine interaction experienced by polarons in PP states at different locations along the interface \cite{35}.  As a simple estimate, assuming that the hydrogen nuclei are distributed evenly in space, the term $E_{HF}^2$ in Eq. (\ref{eq5}) is proportional to $\int d^3r |\psi(r)|^4\propto 1/V$ \cite{19}.  Here $\psi(r)$ is the spatial wavefunction of the electron or hole polaron in a PP state and $V$ is the volume that characterizes its spatial extent.  (From Eq. (\ref{eq5}), $B_{HF}$ is then proportional to $1/\sqrt{V}$.)  Incorporating this effect, the correlation terms in Eq. (\ref{eq5}) have dependence on $V$ that is written as:

\begin{equation}
\label{eq11}
J=\frac{\widetilde{J}}{V}
\end{equation}

$\widetilde{J}$ is a generic quantity that depends on the spatial density of hydrogen nuclei.  For simplicity, in the calculation $\widetilde{J}$ is assumed to be the same for donor and acceptor materials.  Incorporating Eq. (\ref{eq11}) into the model, we can fit the experimental data with appropriate parameters.  $\tau_0=2$ ns and $K_S=(0.6$ ns$)^{-1}$ are used from the previous discussions.  The two parameters that vary among the different curves are $V$ (normalized by the $550\sim600$ nm curve, which we choose to define a reference value $V_{ref}$) and $K_T/K_S$.  The parameter values for all the curves are shown in Table I.  In fitting the data, the $type$ $II$ correlation function is used because it shows better agreement with the experimental data than the $type$ $I$ function.  The model calculation results are shown as lines in Fig. \ref{6}(a).  Lower energies of the emitted photons correspond to more compact states; therefore the polarons interact with fewer nuclei \cite{35}.

\begin{table}
\caption{\label{table1}%
Parameters used in fitting experimental data of Fig. \ref{6}(a)
}
\begin{ruledtabular}
\begin{tabular} {ccc}
\textrm{Wavelength (nm)}&
\textrm{$(V/V_{ref})^{\frac{1}{3}}$\footnote{With respect to the $550\sim600$ nm curve}}&
\textrm{$K_T/K_S$}
\\
\colrule
$450\sim500$ & 1.14 & 1.74 \\
$500\sim550$ & 1.06 & 1.4 \\
$550\sim600$ & 1 & 1.27 \\
$600\sim650$ & 0.95 & 1.23 \\
$650\sim700$ & 0.9 & 1.2 \\
$700\sim750$ & 0.85 & 1.18 \\
\end{tabular}
\end{ruledtabular}
\end{table}

\section{Incorporation of Exchange Coupling in the Polaron Pair States}

\begin{figure}
\centering\includegraphics[width=1.8in]{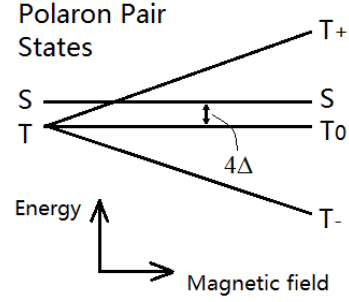}% Here is how to import EPS art
\caption{\label{7} A schematic of the PP energy levels as a function of an external magnetic field, including an energy split due to exchange coupling.}
\end{figure}

Some magneto-luminescence experiments for organic semiconductors reveal a fine structure in MFE for very low magnetic fields ($<2$ mT).  The MFE consists of a slight decrease in luminescence before a substantial increase \cite{11,12,36}.  This observation implies that the effect of hyperfine interaction does not decrease monotonically with increasing magnetic field.  Considering that the magnetic field determines the Zeeman energy and the low-field structure occurs at very weak fields, a reasonable explanation is that a weak exchange interaction in the PP state lifts the degeneracy of the singlet and triplet states, as shown in Fig. \ref{7}.  The small energy difference is due to the relatively weak coupling of the electron and the hole polarons in the PP state.  As the magnetic field increases, the magnitude of the energy difference between $S$ and $T_+$ states decreases initially and increases subsequently.  Therefore the correlation between the two states exhibits an initial increase followed by a decrease. 

The effect of weak exchange coupling in the PP states is readily incorporated into the previous model.  The new Hamiltonian of $0th$ order can be written as \cite{12}:

\begin{equation}
\label{eq12}
\begin{split}
H_0^n & =H_Z^n+H_{PPEC}^n \\
 & =\frac{g\mu_B}{\hbar}\vec B \cdot (\vec S_h^n + \vec S_e^n)-\frac{4}{\hbar^2}\Delta(\vec S_h^n \cdot \vec S_e^n)
\end{split}
\end{equation}

$PPEC$ denotes Polaron Pair Exchange Coupling.  $\Delta$ describes the energy scale of the exchange coupling in PP states.  Using similar derivation procedures \cite{20}, analytical expressions of the PP density matrix are obtained:

\begin{subequations}
\label{eq13}
\begin{gather}
\rho_{22}=\frac{C_1 B_2 - C_2 B_1}{A_1 B_2 - A_2 B_1}R
\\
\rho_{33}=\frac{C_1 A_2 - C_2 A_1}{B_1 A_2 - B_2 A_1}R
\\
\rho_{11}=\frac{2R+J_O^{21}\rho_{22}+J_O^{31}\rho_{33}}{2K_T+J_O^{21}+J_O^{31}}
\\
\rho_{44}=\frac{2R+J_O^{42}\rho_{22}+J_O^{31}\rho_{33}}{2K_T+J_O^{42}+J_O^{31}}
\end{gather}
\end{subequations}

where

\begin{subequations}
\begin{gather}
A_1=\frac{J_O^{21}J_O^{31}(J_O^{21}-J_O^{42})}{4K_T+2J_O^{21}+2J_O^{31}}-\frac{1}{2}J_O^{31}J_O^{21}-\frac{1}{2}J_O^{31}J_O^{42} \nonumber\\
-J_O^{31}J_S^{32}-J_O^{42}J_S^{32}-J_O^{31}K_S \nonumber
\\
B_1=\frac{(J_O^{31})^2(J_O^{21}-J_O^{42})}{4K_T+2J_O^{21}+2J_O^{31}}+J_O^{42}J_O^{31} \nonumber\\
+J_O^{31}J_S^{32}+J_O^{42}J_S^{32}+J_O^{42}K_T \nonumber
\\
C_1=-\frac{J_O^{31}(J_O^{21}-J_O^{42})}{2K_T+J_O^{21}+J_O^{31}}+J_O^{42}-J_O^{31} \nonumber
\end{gather}
\end{subequations}

$A_2$, $B_2$, and $C_2$ are obtained by interchanging $J_O^{21}$ and $J_O^{42}$ in $A_1$, $B_1$, and $C_1$, respectively.  The $J$ terms are defined as:

\begin{subequations}
\begin{gather}
J^{ab}=J^{ab(e)}+J^{ab(h)} \nonumber
\\
J^{ab(e,h)}=\frac{\alpha E_{HF(e,h)}^2}{\hbar^2}\int f(|\tau|/\tau_{0(e,h)})e^{i\tau(E_b-E_a)/\hbar}d\tau \nonumber
\end{gather}
\end{subequations}

The values of $\alpha$ are the same as those in the previous section.  $\alpha=2/3$ for the case of antiparallel spins and $1/3$ for the case of parallel spins.  Note that for convenience the basis has been changed: $1=T_+$, $2=S$, $3=T_0$, and $4=T_-$.  The corresponding energies for $0th$-order Hamiltonian are $E_1=g\mu_B B-\Delta$, $E_2=3\Delta$, $E_3=-\Delta$, and $E_4=-g\mu_B B-\Delta$, where $\Delta=g\mu_B B_{PPEC}$ characterizes the exchange coupling in the PP states.  The exciplex formation rates are given by:

\setcounter{equation}{15}
\begin{subequations}
\label{eq14}
\begin{gather}
\chi_S=L^S\rho_{22}
\\
\chi_{T0}=L^T\rho_{33}
\\
\chi_{T+}=L^T\rho_{11}
\\
\chi_{T-}=L^T\rho_{44}
\end{gather}
\end{subequations}

\begin{figure} 
\centering\includegraphics[width=3.34in]{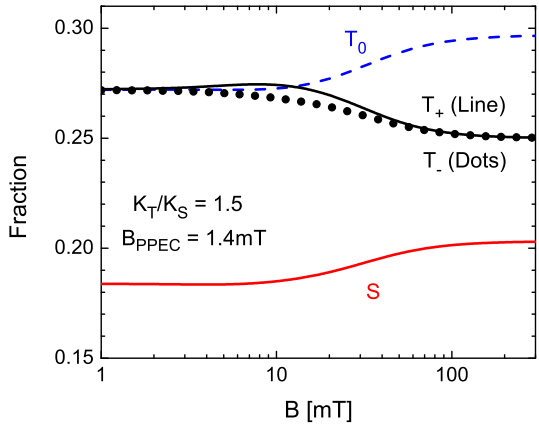}% Here is how to import EPS art
\caption{\label{8} (Color Online) Fractions of singlet and triplet exciplex states as a function of an external magnetic field.  A $1.4$ mT exchange coupling in PP states is included.  The parameters are the same as in Fig. \ref{3}.}
\end{figure}

When $\Delta\rightarrow0$, the combination of Eqs. (\ref{eq13}) and (\ref{eq14}) is identical to Eqs. (\ref{eq4}) and (\ref{eq8}).  The correlation functions have the same forms as in the previous section.  For example, using the $type$ $I$ correlation function, the $J$ terms can be written explicitly as:

\begin{subequations}
\begin{gather}
J_O^{ab(e,h)}=\frac{4E_{HF(e,h)}^2}{3}\frac{\tau_{0(e,h)}}{\hbar^2+\tau_{0(e,h)}^2(E_a-E_b)^2}
\\
J_S^{ab(e,h)}=\frac{2E_{HF(e,h)}^2}{3}\frac{\tau_{0(e,h)}}{\hbar^2+\tau_{0(e,h)}^2(E_a-E_b)^2}
\end{gather}
\end{subequations}

Fig. \ref{8} plots the exciplex fraction as a function of the magnetic field.  The parameters are the same as those in Fig. \ref{3}.  An additional $B_{PPEC}=1.4$ mT ($\Delta\approx0.16$ $\mu$eV) is incorporated into the model.  We observe that there is a peak in the $T_+$ fraction as the magnetic field increases.  This is consistent with the energy schematic in Fig. \ref{7}.  When the $T_+$ and $S$ energy levels cross, the hyperfine interaction between them has a maximum.  When a magnetic field is applied the symmetry between $T_+$ and $T_-$ states is broken and $\chi_{T+}$ is no longer equal to $\chi_{T-}$ (different from Fig. \ref{3}).  When $B\gg B_{PPEC}$, the effect of PP exchange coupling is negligible and their fractions both reach $0.25$.

\begin{figure}
\centering\includegraphics[width=3.34in]{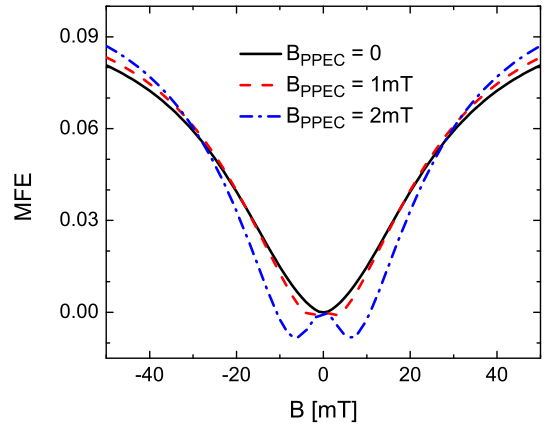}% Here is how to import EPS art
\caption{\label{9} (Color Online) MFE (defined in Eq. (\ref{eq10})) as a function of an external magnetic field.  A variation of the PP exchange coupling is applied to the three curves.  The correlation function used is $type$ $I$.  Other parameters are the same as in Fig. \ref{3}.}
\end{figure}

Fig. \ref{9} shows the MFE for three different exchange interaction strengths in the PP states.  The corresponding energies are $\Delta=0$ , $0.12$ $\mu$eV and $0.23$ $\mu$eV.  The parameters used are the same as in Figs. \ref{3} and \ref{8}.  The result is also consistent with Fig. \ref{7}.  Because of the weak exchange coupling in the polaron pairs, the MFE becomes negative before it turns positive.  When the exchange interaction is greater, the $w$-shape at low field is more significant.

\section{Conclusion}

As a final remark, we comment on different recombination mechanisms in the OLEDs.  As mentioned in section II, a polaron pair may form at the heterojunction interface or in the bulk material, resulting in formation of an exciplex or a bulk exciton.  Usually an exciplex extends at least over two adjacent molecules while an exciton resides on a single molecule.  Hence, the PP state involved in exciton formation is also likely to be more localized than that occurring during exciplex formation.   This in turn implies that the effect of PP exchange coupling is more prominent for the bulk exciton mechanism.  That is a possible explanation that the fine structure is observed in Ref. \cite{9} but not in our experiment in Fig. \ref{6}.  For the type of correlation functions in the system, the experimental observations suggest that devices with exciplex recombination tend to have a $type$ $II$ behavior (Fig. \ref{6}), while devices with exciton recombination are more likely to be in the $type$ $I$ form \cite{37,38}.  But firm conclusions cannot be reached without further explorations.

	We have presented a theoretical model for magneto-electroluminescence in organic semiconductor light emitters.  It yields insight into the physics of the hyperfine interaction and Zeeman effect for polarons in organic molecules.  It also illuminates how singlet/triplet exciplex formation rates and the spatial extent of the polaron pair control the optoelectronic properties.  

\begin{acknowledgments}
This work was supported by a University of Minnesota Doctoral Dissertation Fellowship, the MRSEC program of the National Science Foundation at University of Minnesota under Award Number DMR-0819885, and the LDRD program at Los Alamos National Laboratory.
\end{acknowledgments}

\end{document}